\documentclass[aps,prd,twocolumn,showpacs,amsmath,preprintnumbers,nofootinbib,superscriptaddress,secnumarabic]{revtex4}

\def\scn#1#2{\section{#1}\lb{#2}}
\def\sscn#1#2{\subsection{#1}\lb{#2}}

\usepackage{epsfig,amssymb,amsfonts,verbatim}

\usepackage{graphicx}
\usepackage{color}

\def\bfl{\begin{flushleft}}
\def\efl{\end{flushleft}}
\def\bfr{\begin{flushright}}
\def\efr{\end{flushright}}
\def\bc{\begin{center}}
\def\ec{\end{center}}
\def\be{\begin{equation}}
\def\ee{\end{equation}}
\def\ba{\begin{eqnarray}}
\def\ea{\end{eqnarray}}
\def\baa#1{\begin{array}{#1}}
\def\eaa{\end{array}}
\def\bw{\begin{widetext}}
\def\ew{\end{widetext}}
\def\nn{\nonumber }
\def\lb#1{\label{#1}}
\def\bit{\begin{itemize}}
\def\eit{\end{itemize}}

\def\schrod{Schr\"odinger  }

\newtheorem{Def}{Def.}[section]

\newtheorem{Prp}[Def]{Proposition}

\def\L{\Lambda}

\begin{document}

\preprint{\small Cent. Eur. J. Phys. 11 (2013) 325-335
[arXiv:1204.6380]}

\title{
Singularity-free model of electric charge in physical vacuum: Non-zero spatial extent and mass generation
}

\author{Vladimir Dzhunushaliev}
\email{vdzhunus@krsu.edu.kg}
\affiliation{
Department of Theoretical and Nuclear Physics, 
Kazakh National University, Almaty, 010008, Kazakhstan}
%\affiliation{Institute of Physicotechnical Problems and Material Science of the NAS
%of the
%Kyrgyz Republic, 265 a, Chui Street, Bishkek, 720071,  Kyrgyz Republic}
\affiliation{
Institut f\"ur Physik, 
Universit\"at Oldenburg, Postfach 2503, 
D-26111, Oldenburg, Germany}

\author{Konstantin G. Zloshchastiev}%
\email{k.g.zloschastiev@gmail.com}
\affiliation{
School of Chemistry and Physics,
University of KwaZulu-Natal,
Private Bag X01, Scottsville 3209, Pietermaritzburg, 
South Africa}

%\author{Vladimir V. Dzhunushaliev}
%\affiliation{Institute for Basic Research, Eurasian National University, 
%Astana, 010008, Kazakhstan \\
%and \\
%Institut f\"ur Physik, Universit\"at Oldenburg, Postfach 2503
%D-26111 Oldenburg, Germany }

%\author{Konstantin G. Zloshchastiev}

%\affiliation{
%National Institute for Theoretical Physics (NITheP),
%Stellenbosch Institute for Advanced Study,
%Stellenbosch, South Africa}
%\affiliation{and Institute of Theoretical Physics, University of Stellenbosch, Stellenbosch 7600, South Africa}
%\affiliation{
%Department of Physics and Center for Theoretical Physics,
%University of Witwatersrand,
%Wits 2050, Johannesburg, South Africa}
%\affiliation{Condensed Matter Group, 22 Jln Melur 14, Tmn Melur, Ampang, Selangor DE, Malaysia}

%\date{~ ~~~~~~~~~~~~~~~~~~~~~~}
%\date{~Received: 12 Jan 2001 [LANL] ~}
%\date{~Received: 26 May 2000 [PRL], 1 June 2000 [LANL] ~}
%\date{Received \today}

%\scriptsize%\footnotesize

\begin{abstract}
We propose a model of a spinless electrical charge as a self-consistent
field configuration of the electromagnetic (EM) field interacting with a physical vacuum
effectively described by the logarithmic quantum Bose liquid.
We show that, in contrast to the EM field propagating in a trivial vacuum, a regular solution does exist, 
and both its mass and spatial extent emerge naturally from dynamics.
It is demonstrated that the charge and energy density distribution acquire Gaussian-like form. 
The solution in the logarithmic model is stable and energetically 
favourable, unlike that obtained in a model with a quartic (Higgs-like) potential.
\end{abstract}

\maketitle

%\narrowtext

%\small \newpage

\scn{Introduction}{sec-i}

Two of the oldest actual problems in fundamental particle physics
relate to finite self-energy and the possible extendedness of  electrically charged
elementary particles.
%, such as the electron. 
This research direction is complicated by the fact that no experimental evidence of either the internal structure or spatial extent of, e.g., the electron has been found down to $10^{-16}$ cm. 
The mere postulate that a certain amount of matter with mass, charge and spin can be located inside a set of zero spatial measure looks implausible to a physicist's mind. 
This assumption, however, might be one of the reasons why unphysical divergences appear in quantum field theory (QFT). This difficulty already arises  at the classical level: 
according to the standard theory of electromagnetism, the electrical field of a point charge in completely empty space (trivial vacuum) is described by the inverse square law, therefore the energy density of the electrical field integrated over
the whole space turns out to be infinite. As a result, the total mass-energy of the point charge, together with its field, becomes infinite,
therefore, such a system would be impossible to move.
%which obviously contradicts to our experience.
At the quantum level, this problem manifests itself in ultraviolet
divergences appearing in loop diagrams. In some theories, these divergences can be removed by means of regularization and renormalization procedures.
This can be very useful for doing specific computations, but does not shed much light upon the essence of the problem 
\cite{stern52,stern52a,stern52b,stern52c,stern52d}. 
Theoretical attempts towards better understanding
% the above-mentioned problem
should not, however, be abandoned.

Historically the first effort to address this problem was probably a model which described the electron as a ball with spatially distributed electrical charge. That model conflicted with relativity, however,  because the ball was assumed to be absolutely rigid. The description of spin was also not clear.
Similar difficulties were found in models proposed by Abraham and Lorentz \cite{ablo,abloa,ablob}, although work in that direction continues \cite{Appel:2000th,Appel:2000tha}.
Dirac's shell model of the extended electron, together with its subsequent modifications 
and variations \cite{Dirac:1962iy,Dirac:1962iya,Dirac:1962iyb,Dirac:1962iyc},  provides another notable research direction.  
%Einstein's wormhole approach also icluded extended electrical charge;
Yet another model of an extended electrical charge was the Einstein's wormhole approach.
In his model of an electron
the electrical flux lines enter one side of the wormhole and exit from another, resulting in the front side looking like a negative charge and the rear like a positive charge. 
The wormhole models were criticized by Wheeler for issues of stability, non-quantized charge, wrong mass-charge ratio and spin \cite{wheelbook}. Numerous attempts towards finding regular particle-like solutions were made in  conventional and nonlinear electrodynamics \cite{Born:1934gh}, both with and without engaging general relativity \cite{Bronnikov:1979ex,Bronnikov:1979exa,Finster:1998ws,Datzeff:1980tg,Bonnor:1989iw,Bonnor:1989iwa,Markov:1970hv,Markov:1970hva,Zaslavskii:2010yi,Zaslavskii:2010yia,Zaslavskii:2010yib,Zaslavskii:2010yic}.
Another interesting approach is wave-corpuscle mechanics (for a review see \cite{bf08}).
The general mathematical formalism used therein formally resembles the one used in this paper.
The important difference, however, is about underlying physics: 
the origin and explicit form of their nonlinear self-interacting wave term $G (|\psi|^2)$
are not specified
on physical grounds and a fully satisfactory particle-like solution is not given.
  
A popular approach to the classical electron model was made using the Einstein-Dirac \cite{Bronnikov:1979exa} and Kerr-Newman (KN) solutions \cite{KNold,KNold1,KNold2,KNold3,KNold3a,KNold3b,KNold3c,KNold4,KNold5,KNold5a,KNold6,KNold7,KNold8} 
(an extensive bibliography can be found in \cite{Burinskii:2011id}). 
While the original KN solution does have the correct gyromagnetic ratio, it also
contains a naked singularity and thus 
requires an additional mechanism to circumvent the regularization problem; the story is far from being complete yet. 
%Another popular approach, the string/brane theory, attempts to solve the spatial extent and self-energy problems by postulating extended objects as being fundamental, from the beginning. 
%In this connection a natural question arises of what are the particles such extended objects are supposed to be ``made of'', and whether the theory's set of fundamental degrees of freedom is complete.

Intuitively, from a fundamental theory  one would expect that  spatial extent is not {\it ab initio}
built-in, but naturally
emerges from dynamics.
In this paper we propose a model of a charged particle whose spatial extent,
observable charge and mass emerge as a result of the interaction of the
EM field with the physical vacuum.
For simplicity we neglect internal degrees of freedom,
such as spin, isospin, etc., so that we can assume spherical
symmetry where possible.
The resulting solution describes a charged object 
which does not have a boundary in a classical sense;
its stability is supported not by surface
tension but
by nonlinear quantum effects in the bulk.
This makes our model more realistic 
from the quantum-mechanical point of view, 
since the actual observability of a definite boundary with smooth surface would be as contradictory to
the quantum uncertainty principle as the notion of a smooth trajectory
or worldline
in the quantum realm.
This can be shown by performing a simple \textit{Gedankenexperiment}: making an extended object with a definite
boundary  propagate through space and measuring the 
velocity and position uncertainties on its surface.
In turn, it means that the surface tension is a well-defined notion only in the classical limit,
but for more fundamental purposes it must be used with utmost care.

The structure of the paper is as follows.
The phenomenological approach to physical vacuum is described in the next section,
the main equations of the model can be found in Section \ref{sec-fieq},
the regular solution and its
properties are analyzed, both analytically and numerically, in Section \ref{sec-pls}.
A comparison between our solution  and its Higgs-type (quartic) counterpart is done in Section \ref{sec-vs},
and  conclusions are drawn in Section \ref{sec-con}.

\scn{Physical vacuum}{sec-pvac}

As mentioned above, the Coulomb divergence problem essentially means that one cannot find 
regular particle-like solutions of the Maxwell field in empty space, not even in general relativity \cite{bs76}. From a quantum physicist's point of view, however, this problem is not as severe as it looks to a non-quantum theorist, because the notion of absolutely empty space (or ``mathematical vacuum'') cannot be realized in nature anyway. 
This is because the existence of such space seriously contradicts quantum-mechanical laws. According to the latter, the genuine (physical) vacuum must be a non-trivial quantum medium which acts as a non-removable background and affects particles propagating through \cite{stern52,Latorre:1994cv}.

At this time, no commonly accepted theory of physical vacuum exists.
The amount of experimental and observational data is still too far from conclusively identifying a single model. 
One of the candidate theories lies within the framework of the superfluid vacuum approach \cite{dir51,dir51a,Sinha:1976dw,Sinha:1976dwa,Sinha:1976dwb,vol03}.
This theory
is based on the idea \cite{Zloshchastiev:2009zw,Zloshchastiev:2009zwa,Zloshchastiev:2009aw}
that a physical vacuum can be viewed as some sort of background superfluid condensate
described by the logarithmic wave equation  
(the latter was studied previously on grounds of the dilatation
covariance \cite{ros69}
%, quantum dissipation \cite{kos73} 
or separability \cite{BialynickiBirula:1976zp,BialynickiBirula:1976zpa,BialynickiBirula:1976zpb}).
%, the modern interpretation can be found in the appendix of \cite{az11}). 
%The elementary constituent of this superfluid is unknown: it can be the yet undiscovered particle but this condensate can also be a giant collective state of helium - considering the abundance of this element in the Universe and that its excitation spectrum is known to interpolate between the relativistic and non-relativistic regimes \cite{az11}.
It was shown that small fluctuations of the logarithmic condensate obey the Lorentz symmetry and can be interpreted not only as the relativistic particle-like  states but also as the gravitational ones \cite{Zloshchastiev:2009aw} depending on a type of mode.
In this approach, therefore,  the Lorentz symmetry is not an exact symmetry of nature,
but rather pertains to small fluctuations of the physical vacuum,
and thus gets deformed at high energies and/or momenta.

As long as the superfluid vacuum approach must be fully consistent and applicable to reality,
one would expect that if the empty space is replaced by the logarithmic vacuum condensate then the behaviour of the
conventional Maxwell field becomes regular in the  presence of 
particle-like solutions. 
This behaviour is going to be the main subject of the current study.
One would also expect that the spatial extent mentioned above must
appear naturally in the approach. This has already been shown at the non-relativistic level in \cite{az11}, so in the current study we will investigate the relativistic case.

\begin{figure}[htbt]
\begin{center}
	\epsfig{figure=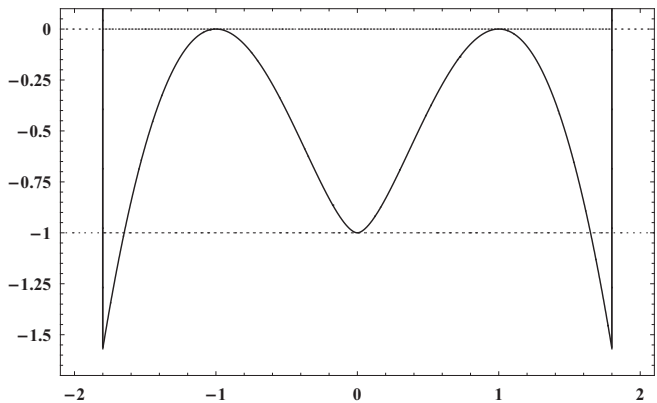,width=  1.01\columnwidth}\end{center}
	\caption{
	The field-theoretical potential \eqref{e-vstart} (in units of $\breve{\beta} \breve{a}^3$)
versus $\text{Re}(\Psi)$ (in units of $\breve{a}^{-3/2}$). Vertical lines
conditionally represent inequality (\ref{e-finitpsi}). 
In an approximation when the symmetry-breaking energy scale is much less
than the vacuum one ($\gtrsim 10$ TeV) these walls can be assumed infinite.
	}
\label{f:ftpot}
\end{figure}

The effective low-energy Lagrangian for the Maxwell field interacting with small fluctuations of the physical vacuum was proposed in \cite{Zloshchastiev:2009aw}, along with a mass generation mechanism
which was analogous to the Higgs one.
In this paper we propose a different mass generation mechanism
which uses the same Lagrangian, except that the scalar potential is assumed to be ``upended''.
As will be shown below, the latter solves the issue of a wrong sign in the quadratic (mass) term
of the  potential at energies above the symmetry-breaking scale - when
symmetry is unbroken and false vacuum is stable. 
In the case of  three spatial dimensions the action is proportional to $\int\!d^4 x\, {\cal L}$ where, adopting the natural units $c=\hbar = \epsilon_0 =1$ and metric signature $(+---)$, we assume
\begin{equation}
{\cal L} = -\frac{1}{4} F_{\mu \nu} F^{\mu \nu} + \frac{\breve{a}}{2}  \left|
		D_\mu \Psi
	\right|^2 - 
V( |\Psi |^2) 
,
\label{e-lagstart}
\end{equation}
where $D_\mu  = \partial_\mu  + i g A_\mu $,
$F_{\mu\nu} = \partial_\mu A_\nu - \partial_\nu A_\mu$,
and the vacuum-induced field potential 
%\textcolor{red}{ $V(\psi) = - \frac{1}{\beta} |\psi|^2 \ln \left( a^3 |\psi|^2 \right)$. }
is defined as (up to an additive constant)
\be\label{e-vstart}
	V( |\Psi |^2) 
	= - 
	\breve{\beta}^{-1}
	\left\{
	 |\Psi|^2
	\left[
	\ln{(\breve{a}^3 |\Psi|^2)}
	-1
	\right]
	+ \breve{a}^{-3}
	%+ \beta {\cal V}_0
	\right\}	
	,
\ee
where $\breve{a}$ and $\breve{\beta}$ are parameters of  dimensionality length in adopted units.
In the underlying theory of superfluid vacuum the parameter $\breve{\beta}$ is related
to the quantum (non-thermal)  temperature which is conjugated
to the Everett-Hirschmann information entropy
% which can be interpreted a measure of the wafefunctions spreading 
\cite{az11} whereas
  $\breve{a}$ can be related to the characteristic inhomogeneity scale of the superfluid \cite{zlo2012}.
The potential (\ref{e-vstart}) is regular in the origin - while the logarithm itself diverges there, the factor $|\Psi|^2$  recovers regularity. 
There is a local maximum at
$
|\Psi|_\text{max} = \breve{a}^{-3/2}
,
$
i.e.,
it always has the (upside-down) Mexican-hat shape if plotted as a function of $\Psi$, see Fig. \ref{f:ftpot}. 
In what follows we call this potential \textit{logarithmic} - due to the property $d V / d |\Psi|^2 \propto \ln{( \breve{a}^3 |\Psi|^2)}$
which yields the logarithmic term in the corresponding field equation.

We emphasize that, according to this approach, the Lagrangian (\ref{e-lagstart}) 
is an approximate one, thus it is not valid for arbitrarily large (or short) scales of energy (or length) which means that the $\Psi$ field cannot take arbitrarily large values:
\be\lb{e-finitpsi}
|\Psi| \leqslant |\Psi_c| < \infty
,
\ee
where $|\Psi_c| = \lim\limits_{E\to E_0}|\Psi|$ is some limit value
corresponding to the cutoff energy scale $E_0$ which is also the characteristic energy scale of the vacuum. 
In the effective theory the appearance of upper bound for $|\Psi |$ will
be shown in Section \ref{sec-appr}
below,
whereas
in a full theory it
could result from, for instance, the normalization condition for $\Psi$,
similarly to the one in the theory of Bose-Einstein condensation.
The constraint \eqref{e-finitpsi} 
also means that the potential \eqref{e-vstart} does not have to be bounded from below, as in a standard relativistic QFT. Alternatively, one can take the potential \eqref{e-vstart} with an opposite sign,  thus making it bounded from below at positive $\breve{\beta}$, but treat $\Psi$ as a phantom field.

Further,
performing the rescaling
\be
\psi = \sqrt{\breve{a}} \Psi, \
%\beta^{-1} = \frac{\breve{\beta}^{-1}}{\breve{a}}, \
\beta = \breve{a} \breve{\beta}, \
a = \breve{a}^{2/3}
,
\ee
we can rewrite \eqref{e-lagstart} and \eqref{e-vstart} in a more regular form:
\ba
&&
{\cal L} = -\frac{1}{4} F_{\mu \nu} F^{\mu \nu} + \frac{1}{2} \left|
		D_\mu \psi
	\right|^2 - 
V( \psi ) 
,
\label{2-10}\\&&	V( \psi ) 
	= - 
	\beta^{-1}
	\left\{
	 |\psi|^2
	\left[
	\ln{(a^3 |\psi|^2)}
	-1
	\right]
	+ \frac{1}{a^3}
	%+ \beta {\cal V}_0
	\right\}	
.
\lb{e-pot1}
\ea
Note that by expanding the potential in the vicinity of $|\psi|^2= \varepsilon / a^{3}$, $\varepsilon$ being
a non-negative dimensionless
number, 
%to a quadratic order w.r.t. $|\psi|^2$, 
one arrives at 
the following
perturbative expression (up to an additive constant):
\be\label{e-pot1q}
	V( \psi) 
% \approx \frac{\lambda_\text{eff}}{4!} \left( |\psi|^2 - a^{-3} \right)^2 + ... 
\approx 
%\frac{1}{2}
%\frac{\breve{a}^3}{2 \breve{\beta}} 
\frac{\lambda_\text{eff}}{4!}|\psi|^4
+
\frac{1}{2}
m_b^2 |\psi|^2
+ {\cal O} \left((|\psi|^2-\varepsilon / a^{3})^3\right)
,
\ee
where 
$\lambda_\text{eff} = - 12 a^3 / \varepsilon \beta$ is the effective quartic coupling
and $m_b = \sqrt{2 (1 - \ln\varepsilon)/ \beta}$.
If the radicand of $m_b$ is non-negative then
$m_b$ can be interpreted
as the  mass of an effective scalar particle (before the symmetry breaking).
Indeed, one can always quantize the approximate model by analogy with a quartic (hence renormalizable)
scalar QFT in the 
vicinity of a non-trivial vacuum represented by the ground-state solution of the original model (\ref{e-pot1})
which will be discussed in the following sections.
To date, a number of
different quantization approaches have been developed for such cases
- see, e.g., works \cite{Rajaraman:1982is,Zloshchastiev:2000,Zloshchastiev:2000a}
and references therein.

We have now expressed the physical parameters of the scalar sector, such as
mass and coupling, in terms of the primary parameters of our theory. 
If the value of $\varepsilon$ is close to one (or, at least, less than the
base of natural logarithm) then
it is indeed important that the potential (\ref{e-pot1})
has the upside-down Mexican hat shape (cf. $\breve a > 0$ and $\breve\beta > 0$)
otherwise the quadratic term would appear with a wrong sign.
Also  the effective quartic coupling turns out to be negative
in this case which is a remarkable difference from the standard Higgs potential
and thus it can serve for experimental testing.
%The latter issue is known to happen in the scalar sector of the Glashow-Weinberg-Salam theory where it is not explained, to our best knowledge.
In any case, the interpretation based on (\ref{e-pot1q}) is only approximate
(for instance, such series expansion does not converge to (\ref{e-pot1})  for very small $|\psi |$), therefore,
in what follows we will be working with the exact expression for $V$.

\scn{Field Equations}{sec-fieq}

The  field equations corresponding to the Lagrangian \eqref{2-10}
are given by
\be
		\partial_\mu F^{\mu \nu} = j^\nu,
\label{2-20}	
\ee
\be
		D_\mu D^\mu \psi + \frac{\partial \, V}{\partial \psi^*}
	= 
	\left[
	D_\mu D^\mu + \beta^{-1} \ln{(a^3 |\psi|^2)}
	\right] \psi
	=
	0
	,
\label{2-30}
\ee
where the current $j^\nu$ is 
defined as
\be
	j^\nu =  i g \left[
		\left( D^\mu\psi \right)^* \psi - \psi^* \left( D^\mu\psi \right)
	\right]
	. 
\ee
We look for a solution in the electrostatic form 
\be
	A_\mu = \left[ \phi(r), \vec 0 \right], 
	\ 
%\label{2-60}\\
	\psi(t,r) = \text{e}^{-iEt} \psi(r) 
	,
\label{2-70}
\ee
with $E$ being a real-valued constant.
Then the equations of motion become simply
\begin{eqnarray}
	\psi'' + \frac{2}{r} \psi' &=& - \left (
	E - g \phi 
	\right )^2 \psi - 
	\beta^{-1} \psi \ln (a^3 \psi^2),
\label{2-80}\\
	\phi'' + \frac{2}{r} \phi' &=& -2 g \left(
		E - g \phi
	\right) \psi^2 
,
\label{2-90}
\end{eqnarray}
where primes indicate derivatives with respect to $r$. 
Introducing the quantities
\ba
&&
x=	\frac{r}{\sqrt{\beta}}, \
\tilde \psi = a^{3/2} \psi = \breve{a}^{3/2} \Psi, \
\tilde \phi =	\nn\\&&
\qquad
= -g \phi \sqrt{\beta}  = -g \phi \sqrt{ \breve{a} \breve{\beta}} 
, \\&&
\tilde E =	E \sqrt{\beta}  = E \sqrt{ \breve{a} \breve{\beta}} , \
\tilde g = g	\sqrt{\frac{ \beta}{a^3}}  = g\sqrt{\frac{\breve{\beta}}{\breve{a}}} 
,
\ea
we obtain 
\begin{eqnarray}
	\tilde \psi'' + \frac{2}{x} \tilde \psi' &=& - \left (
	\tilde E + \tilde \phi  
	\right)^2 \tilde \psi - \tilde \psi \ln (\tilde \psi^2),
\label{2-110}\\
	\tilde \phi'' + \frac{2}{x} \tilde \phi' &=& 2 
	\tilde g^2 \left(
		\tilde E + \tilde \phi
	\right) \tilde \psi^2 
	,
\label{2-120}
\end{eqnarray}
where primes  indicate derivatives with respect to $x$ when applied to tilded
quantities. 
One can see that equations depend on only one parameter  $\tilde g$, whereas $\tilde E = \tilde E (\tilde g)$ 
can be treated as an eigenvalue at a given $\tilde g$. 
The full theory of physical vacuum would provide the value of $\tilde E$
(or, equivalently, $\tilde \psi (0)$, see the analytical solution section below)
as a function of the primary parameters
(from, e.g., some sort of  normalization condition, cf. \cite{az11}) but
in the approximate theory $\tilde E$ stays a free parameter which
can be fixed only from external considerations.

Further, for a given solution,
the energy density $\epsilon$ is defined as
\begin{equation}
 \epsilon = \frac{1}{2} D_0\psi^* D^0\psi + \frac{1}{2} D_i\psi^* D^i\psi + 
 V(\psi) + \frac{1}{2} |\vec {\mathcal E}|^2
 ,
\label{2-140}
\end{equation}
where $i=1,2,3$
and $\mathcal E = - \nabla \phi$ is the electric field strength. 
Substituting the ansatz \eqref{2-70} we obtain
\begin{equation}
 \beta a^3 \epsilon = \frac{1}{2} 
 \left( \tilde E + \tilde \phi \right)^2 \tilde \psi^2 + 
 \frac{1}{2} {\tilde{\psi'}}^2
  - 
 \tilde \psi^2 \ln \left( \tilde \psi^2 \right) 
 + 
 \frac{1}{2 \tilde g^2}
 \tilde{\phi'}^2
 .
\label{2-150}
\end{equation}
The total energy can be calculated as 
$ W = 
 4 \pi \int \limits_0^\infty r^2 \epsilon(r) dr
 $
so we obtain
\be
W 
 = 
\frac{4 \pi \sqrt{\beta}}{ a^3} \tilde W (\tilde g)
,
\label{2-160}
\ee
where we denoted the dimensionless total energy
\bw
\be\lb{e-totendl}
\tilde W (\tilde g)
=
\frac{1}{2} 
\int \limits_0^\infty 
 \left[
 	\left( \tilde E + \tilde \phi \right)^2 \tilde \psi^2 + 
{\tilde{\psi'}}^2 - 
2 \tilde \psi^2 \ln \left( \tilde \psi^2 \right) + 
\left(\frac{\tilde{\phi'}}{\tilde g} 
\right)^{2} 
 \right] x^2  d x 
.
\ee
\ew
For an observer in the reference frame associated with the center
of mass of a localized solution the quantity $W/c^2$ is equivalent to
the rest mast of a corresponding particle.

\scn{Particle-like solution and its properties}{sec-pls}

\sscn{Asymptotic behaviour}{sec-asymp}

To 
search for the regular solution, the functions $\tilde \psi(x)$ and $\tilde \phi(x)$ should have the
following behaviour near the origin: 
\begin{eqnarray}
	\tilde \psi(x) & = & \tilde \psi_0  + 
	\tilde \psi_2 \frac{x^2}{2} + \mathcal O(x^4), 
\label{4a-10}\\
	\tilde \phi(x) & = & \tilde \phi_0  + 
	\tilde \phi_2 \frac{x^2}{2} + \mathcal O(x^4). 
\label{4a-20}
\end{eqnarray}
When substituting this into \eqref{2-110} and \eqref{2-120} we obtain the solution 
\begin{eqnarray}
	\tilde \psi_2 & = & - \frac{\tilde \psi_0}{3}\left[
		\left( \tilde E + \tilde \phi_0 \right)^2 + \ln \tilde \psi_0^2
	\right],
\label{4a-30}\\
	\tilde \phi_2 & = & \frac{2}{3} \tilde g^2 \left(  
		\tilde E + \tilde \phi_0
	\right) \tilde \psi_0^2 . 
\label{4a-40}
\end{eqnarray}
Further,
the asymptotic behaviour at $x \rightarrow \infty$ is given by 
\begin{eqnarray}
	\tilde \psi \to  \text{e}^{\frac{1}{2}(3- \tilde E^2 - x^2)}
\left[1 
+
{\cal O} (\tilde g^2)
\right]
, \
%\label{4a-50}\\&&
	\tilde \phi \to 
-\frac{\tilde q}{x} 
,
\label{4a-60}
\end{eqnarray}
where $\tilde q$ is some constant to be determined. 
It is instructive to relate the bare charge $g$  to the observable one $q = \tilde q / g$. 
By integrating \eqref{2-90} we obtain
\begin{equation}
	r^2 \mathcal E = 2g \int \limits_0^r \left(
		E - g \phi
	\right) \psi^2 r^2  dr 
,
\label{4a-70}
\end{equation}
and in the limit $r \rightarrow \infty$ 
%according to Eq. \eqref{4a-60} 
we have $r^2 \mathcal E \rightarrow q$ hence 
\begin{equation}
	q = 2g \int \limits_0^\infty  \left(
		E - g \phi
	\right) \psi^2 r^2  d r
.
\label{4a-80}
\end{equation}
Thus, we arrive at the following relation between the bare and observable charges 
\begin{equation}
	q = \frac{2g \beta}{a^3} I (\tilde g )= 
 2 \tilde g \sqrt \frac{\beta}{a^3} I (\tilde g )
 ,
\label{4a-90}
\end{equation}
where we denoted
%the integral $I (\tilde g )$ is equal to  
\begin{equation}
	I (\tilde g )= \int \limits_0^\infty x^2 \left(
		\tilde E + \tilde \phi
	\right) \tilde \psi^2 dx
.
\label{4a-100}
\end{equation}
Expression \eqref{4a-60} shows us that at large distance
we recover the Coulomb potential while the field $\tilde \psi$ decreases exponentially.
Thus, the field $\tilde \psi$ is in fact unobservable, unless very short length scales are probed. 
From the asymptotics of the solution one can infer that the charge radius of the solution
is determined by the parameter $\beta$:
\be
\text{size} \sim \sqrt{\beta}
 \sim \sqrt{\breve{a} \breve\beta}
 ,
\ee
which essentially means that the combination of
parameters 
$\beta = \breve{a} \breve\beta$
must have
an extremely small value for the known elementary particles.
For instance, if
one takes  the values of the classical
radius $e^2/m$ as conservative estimates,
then for
the electron and muon one would obtain constraints of
$
\beta_{(e)} < 10^{-26}\ \text{cm}^2
$
and 
$
\beta_{(\mu)} < 10^{-32}\ \text{cm}^2
,
$
although it is not entirely clear whether the classical radius should be
analogous
to the ``smearing'' size which is
our definition of size here.

\sscn{Approximate analytical solution}{sec-appr}

While the  exact expression for a full analytical solution is unknown, 
it is possible to solve the system \eqref{2-110} and \eqref{2-120} using the approximation of weak EM coupling 
\be\lb{e-gsmall}
\tilde g^2 \ll 1
,
\ee
which is equivalent to 
$g^2 \ll a^3/\beta$
or
$g^2 \ll \breve{a}/\breve{\beta}$.
The 
observational constraints 
suggest that
this approximation might
have a good chance to be valid for the known elementary
particles - unless $\breve a$ turns out to be very small.
On the other hand, as long as our approach is an effective one it has certain 
applicability conditions - and one of them is that the vacuum effects predominate
the electromagnetic ones.
Therefore, large values of $\tilde g $ might push
our approach outside its applicability range and
thus the corresponding approximation is not very interesting
from the physical point of view.

Thus, imposing the boundary conditions
\be\lb{e-bc}
\tilde\phi (0) < \infty, \
\tilde\phi (+\infty) = 0, \
\tilde\psi (+\infty) < \infty, \
\tilde\psi' (0) =0, 
\ee
one obtains a solution which is regular for 
$0 \leqslant r \leqslant +\infty$ (see appendix for the details of derivation):
\bw
\ba
&&
\tilde\phi 
=
-\frac{1}{2 x}
\sqrt{\pi}
\tilde g^2
\tilde E
\, \text{e}^{3 - \tilde E^2}
\text{erf} (x)
+
{\cal O} (\tilde g^4)
,\lb{e-asolphi}\\&&
\tilde\psi 
=
 \text{e}^{\frac{1}{2}(3 - \tilde E^2 - x^2)}
\left\{
1+
\frac{1}{4}
\tilde g^2
\tilde E^2
 \text{e}^{3 - \tilde E^2 - x^2}
\left[
1
+
\frac{\sqrt{\pi}}{2 x}
\left(
2 x^2 + 1
\right)
 \text{e}^{x^2}
\text{erf} (x)
\right]
\right\}
+
{\cal O} (\tilde g^4)
,
\lb{e-asolpsi}
\ea
%\ew
where dimensionless energy
$\tilde E$ can be also expressed via the boundary value
\be
\tilde E
=  \sqrt{3 - \ln{(\tilde\psi_0^2)}}
\left(
1 -
\frac{1}{2} 
\tilde g^2
\tilde\psi_0^2
\right)
+
{\cal O} (\tilde g^4)
,
\ee
with the square root being defined up to a sign.
Of course, this formula is valid only
if the magnitude of $\tilde\psi$ is bounded
from above:
\be
|{\tilde\psi}|
\leqslant
|{\tilde\psi}_0| \leqslant \text{e}^{3/2}
,
\ee
which 
\textit{a posteriori}
affirms
%is yet another manifestation of 
the condition of applicability \eqref{e-finitpsi},
although this upper bound 
does not necessarily saturate the critical value there:
$|\Psi_c| \geqslant (\text{e}/ \breve a)^{3/2}$.

Further, one can check that the electric part indeed has the Coulomb behaviour at large $r$, and then the effective charge can be computed as
\be\lb{e-asolchat}
\tilde q 
=
\frac{1}{2}
\sqrt{\pi}
\tilde g^2
\tilde E
\, \text{e}^{3 - \tilde E^2}
+
{\cal O} (\tilde g^4)
,
\ee
therefore,
the observable charge,
\be\lb{e-asolcha}
q 
=
\tilde q / g
\approx
\frac{1}{2 a^{3/2}}
\sqrt{\pi \beta}
\tilde E
\, \text{e}^{3 - \tilde E^2}
\tilde g
\approx
\frac{1}{2 \breve{a}}
\sqrt{\pi}
\breve{\beta}
\tilde E
\, \text{e}^{3 - \tilde E^2}
g
,
\ee
depends 
%not only on the bare one but
on the whole combination of parameters
describing the interaction of the electromagnetic
field with a physical vacuum.
The dimensionless total energy \eqref{e-totendl} 
turns out to be
%\bw
\be
\tilde W 
=
\frac{3}{16}
\sqrt{\pi}
\, \text{e}^{3 - \tilde E^2}
\left[
2 \tilde E^2 -1
+
\frac{\tilde g^2}{3 \sqrt{2}}
\tilde E^2
\left(
12 \tilde E^2 - 13
\right)
 \text{e}^{3 - \tilde E^2}
\right]
+
{\cal O} (\tilde g^4)
,
\ee
\ew
which can also be written in terms of the observable
charge $q$ and rest mass $W$:
\be\lb{e-massf}
W 
\approx
W_{(0)}
\left[
1+
4\sqrt{2}
\beta^{-1} 
a^3 q^2
\text{e}^{\tilde E^2 -3}
\tfrac{
\tilde E^2 - 13/12}{
\tilde E^2 - 1/2
}
\right]
%+ {\cal O} (\tilde g^4)
,
\ee
where
\be
W_{(0)}
=
\frac{3 }{2}
\frac{\pi^{3/2} \sqrt{\beta}}{a^3}
\left(
\tilde E^2 - \frac{1}{2}
\right)
\text{e}^{3 - \tilde E^2}
,
\ee
so
one can see that the obtained formula does not contain any divergences.

It is also apparent that mass $W$ does not vanish when charge is set to zero, 
which indicates that the theory is also capable of incorporating  
non-charged particles into the scheme, by taking the corresponding limit.
In fact, the mass formula (\ref{e-massf}) implies
that for an electrically charged particle with mass $W$
there can exist not only an antiparticle of the same mass
but also a neutral particle of related mass $W_{(0)}$.
It is interesting that the ratio $W / W_{(0)}$ grows exponentially
with growing $|\tilde E |$, which 
results in two possible scenarios:
(i)
the mass of a neutral partner is very small (yet non-zero)
as compared
to the mass of a charged one: this happens if
$|\tilde E | \gg 1$ (or, equivalently, $|\tilde \psi_0| \ll 1$);
(ii) if $|\tilde E |$ is of order one or less then both masses would be of the same
order of magnitude.
The possible phenomenological implications of this mechanism are discussed
in the conclusion.

\begin{figure}[htbt]
\begin{center}
	\epsfig{figure=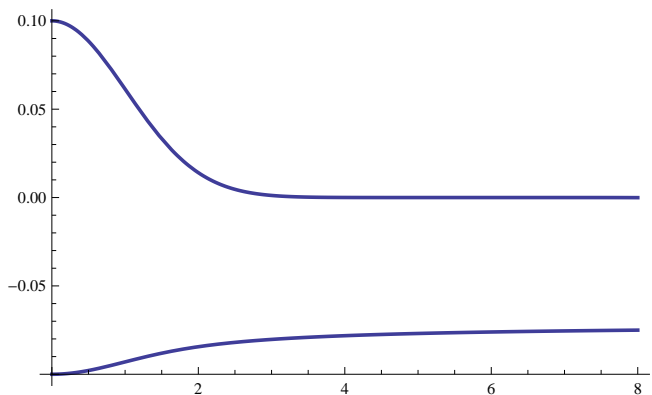,width=  1.0\columnwidth}\end{center}
  \caption{Profiles of $\tilde \psi(x)$ (top curve) and 
  the electrostatic 	potential $\tilde \phi(x)$ (bottom curve),
  computed for 
  eigenvalue $\tilde E = 2.8436935588$.}
 \label{fig1}
\end{figure}

\sscn{Numerical solution and stability}{sec-sol}

\begin{figure}[htbt]
\begin{center}
	\epsfig{figure=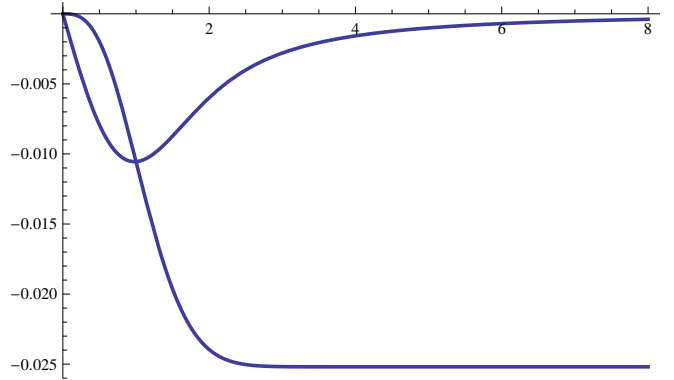,width=  1.0\columnwidth}\end{center}
	\caption{Profiles of the electric field $\tilde{\mathcal E}(x)$ (top) 
	 and $x^2 \tilde{\mathcal E}(x)$ (bottom curve). 
}
\label{fig2}
\end{figure}

\begin{figure}[htbt]
\begin{center}
	\epsfig{figure=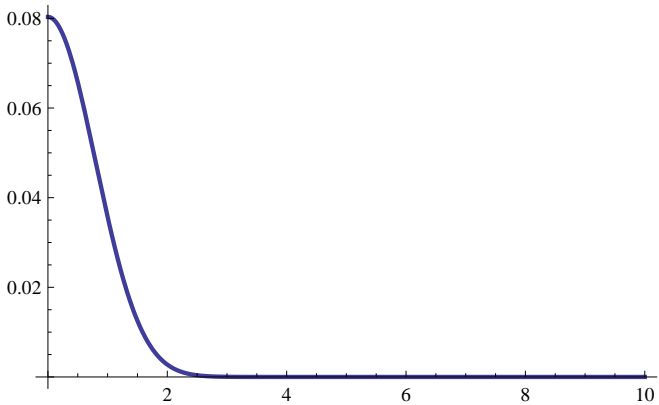,width=  1.0\columnwidth}\end{center}
	\caption{Profile of dimensionless energy density 
	$\beta a^3 \epsilon(x)$, using the same $\tilde E$ eigenvalue as in previous figures.}
\label{fig3}
\end{figure}

While we have managed to find the approximate regular solution analytically, it is important to check that a
regular solution exists for non-small 
$\tilde g$'s and that terms with higher-order powers of $\tilde g$ will not introduce any spatial singularities. For this purpose we solve equations \eqref{2-110} and \eqref{2-120} numerically. For the computations we choose $\tilde g=1$ and the following boundary conditions 
\begin{equation}
 \tilde \psi(0) = 0.1, \quad \tilde \psi'(0) = 0, \quad 
 \tilde \phi(0) = -0.1,  \quad  \tilde\phi'(0) = 0
 ,
\label{2-130}
\end{equation}
and $\tilde E$ is treated as an eigenvalue.
It should be noted that $\tilde \phi(x)$ must be always taken as non-positive on the
positive semi-axis of $x$, due to the 
%way how it was defined and 
asymptotic requirements \eqref{4a-60}.
The numerical solution is presented in Fig. \ref{fig1}, and in Fig. \ref{fig2} the corresponding 
profiles of the electric field $\tilde{\mathcal E} = - d\tilde \phi/dx$ and $x^2 \tilde{\mathcal E}$ are given. 
From these one sees that the electric field is regular at the origin
and asymptotically displays Coulomb behaviour. 
The profile of the dimensionless energy density is shown in Fig. \ref{fig3}. 

%\subsection{Stability}

The direct stability analysis of the solution is complicated by the fact that the perturbed electric field becomes time-dependent, which leads to the appearance of a magnetic field such that this system cannot be regarded as spherically symmetric anymore. It is, however, still possible to use the energy-based arguments as well as to investigate the behaviour of an effective Schroedinger equation potential.
Let us consider the dimensionless total energy $\tilde W$ given by \eqref{e-totendl},
as well as the energy of the field $\tilde \psi$ alone,
$
\tilde W_\psi = \tilde W|_{\phi \to 0}
$,
and the energy of the electric field 
%$\mathcal E = - \tilde \phi'$ 
alone,
$
\tilde W_\phi = \tilde W|_{\psi \to 0}
$.
Then we can define the 
dimensionless binding energy as
\begin{equation}
\Delta \tilde W = \tilde W - (\tilde W_\phi + \tilde W_\psi)
=
 	\frac{1}{2}
\int \limits_0^\infty 
\tilde \phi 
 \left(\tilde \phi + 2  \tilde E \right)  \tilde \psi^2 x^2
 d x
,
\label{4c-40}
\end{equation}
where
% the potential energy $V(\tilde \psi)$ is given by \eqref{e-pot1},
all the potentials are assumed to be given by our regular solution.
One considers the following two possibilities. If binding energy is positive then it is necessary to add a certain amount of energy to create a regular electric charge when coupling to the $\psi$ field. In the opposite case  binding energy gets released during the process, and the energy of the whole configuration is smaller than the sum of the energies of the separate electric and scalar fields. 

Evaluating the binding energy on approximate solution \eqref{e-asolphi}, \eqref{e-asolpsi},
we find that it is negative-definite 
\be
\Delta \tilde W
=
-
2^{-5/2}
\sqrt{\pi}
\left(
\tilde g
\tilde E
\, \text{e}^{3 - \tilde E^2}
\right)^2
+
{\cal O} (\tilde g^4)
,
\ee
which means that 
the creation of the regular electric charge in the logarithmic model  is energetically favourable.

\begin{figure}[htbt]
\begin{center}	
	\epsfig{figure=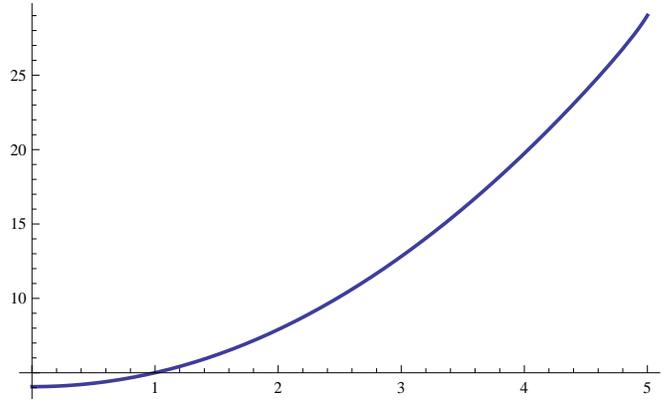,
	width=1.0\columnwidth}
	\end{center}
	\caption{Effective potential \eqref{e-efflog} versus $x$ at $\tilde g$=1, using the same $\tilde E$ eigenvalue as in previous figures.}
\label{f-efflog}
\end{figure}

Another way to study the stability of the solution is to write it as a solution
of the \schrod equation for a fictitious particle 
\be
	- \Delta \Psi + V_\text{eff} (x) \Psi =  \varepsilon \Psi , 
\label{4c-50}
\ee
where $\Delta = d^2/dx^2$, $\varepsilon = \tilde E^2$
and the effective potential is derived as
\be\lb{e-efflog}
	V_\text{eff} (x) = -2 \tilde E \tilde \phi(x) - 
	\tilde \phi^2(x) - \ln{[\tilde \psi^2(x)]}
	,
\ee
where 
the tilded potentials are given by our regular solution.
According to \eqref{4a-60}, the asymptotic behaviour of the solution 
implies that the effective potential $V_\text{eff}  \propto x^2$ at large $x$, (see 
also Fig. \ref{f-efflog}),
and thus the ``particle'' is always localized in a finite region of $x$.
With respect to the solution itself this means that it cannot
spread or be destroyed when subjected to small perturbations. 

%\sscn{Solution interpretation}{s-intp}

\begin{figure}[htbt]
\begin{center}
	\epsfig{figure=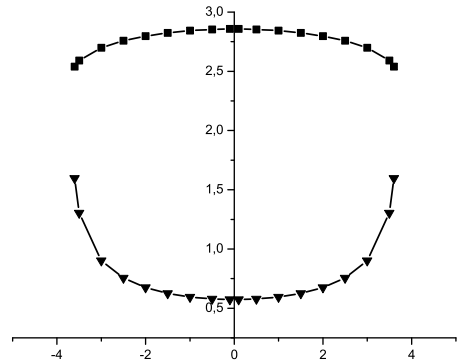,width=  1.0\columnwidth}\end{center}
	\caption{The profiles of the dimensionless energy 
	$\tilde W(\tilde g)$ ($\blacktriangledown$-curve) and 
	the eigenvalues $\tilde E(\tilde g)$ ($\blacksquare$ - curve).}
\label{I_w_phi_0}
\end{figure}

\scn{Logarithmic versus quartic potential}{sec-vs}

One may wonder whether the singularity-free solution exists when the scalar sector of our model is controlled not by the logarithmic potential \eqref{e-pot1} but by a more orthodox one, such as the Higgs-type (quartic) potential:
%\footnote{We remind that the Derrick theorem forbids the regular solutions' existence in  the direct Higgs potential}
\begin{equation}
 V_H(\psi) = - \frac{\varkappa}{4} \psi^4 + \frac{m^2}{2}\psi^2.
\label{2c-10}
\end{equation}

\begin{figure}[htbt]
	\begin{center}
		\epsfig{figure=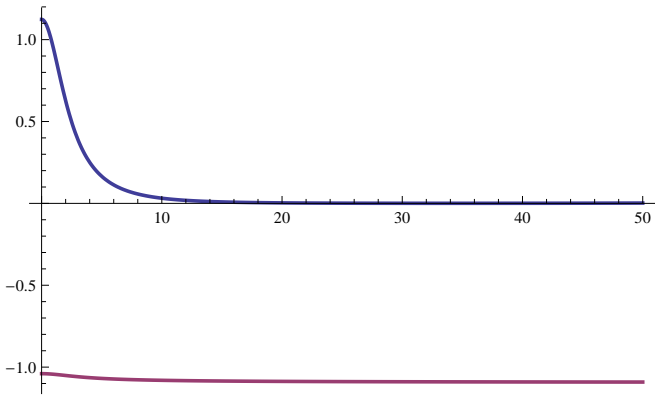,width=  1.0\columnwidth}
	\end{center}
	\caption{The profiles of $\tilde \psi(x)$ (top) and $\tilde \phi(x)$ (bottom)
for the quartic model. 
	Eigenvalue 
	$\tilde \psi(0) = 1.12345$, $\tilde g=0.1$ and parameters 
	$\tilde E=0.1, \lambda=1.0$ have been used.}
\label{higgs_functions}
\end{figure}

\begin{figure}[htbt]
	\begin{center}
		\epsfig{figure=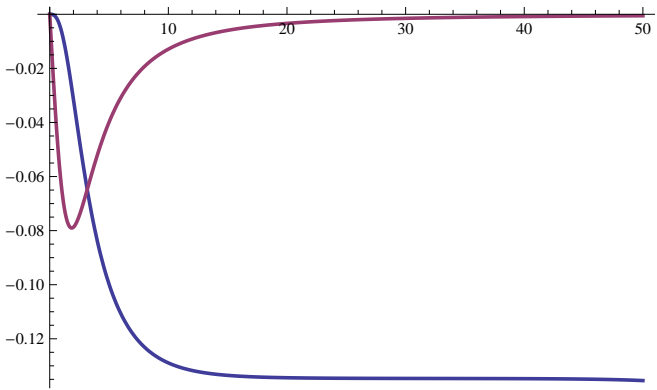,width=  1.0\columnwidth}
	\end{center}
	\caption{Electric field versus $x$ for the quartic model. The top curve is  
	$10 \, \tilde{\mathcal E}(x)$, the bottom one is 
	$x^2 \tilde{\mathcal E}(x)$.
	}
\label{higgs_electric_field}
\end{figure}

Corresponding dimensionless equations with the ansatz  \eqref{2-70} are 
\begin{eqnarray}
	\tilde \psi'' + \frac{2}{x} \tilde \psi' &=& \left[ \left (
	\tilde E + \tilde \phi 
	\right )^2 - \lambda \tilde \psi^2 + 1 \right] \tilde \psi , 
\label{2c-20} \\ 
	\tilde \phi'' + \frac{2}{x} \tilde \phi' &=& 2 \tilde g^2 \left(
		\tilde E + \tilde \phi
	\right) \tilde \psi^2 . 
\label{2c-30}
\end{eqnarray}
where we introduced following dimensionless quantities 
\begin{equation}
 x = m r, \  \tilde \psi = \frac{\psi}{\psi(0)}, 
\ \tilde E = \frac{E}{m}, 
 \ \tilde \phi = -\frac{g \phi}{m}, \ 
 \lambda = \frac{\psi(0)^2 \varkappa}{m^2}. 
\end{equation}
The boundary conditions are 
\begin{equation}
 \tilde \psi'(0) = 0; \quad 
 \tilde \phi(0) = -1.04;  \quad \tilde \phi'(0) = 0.
\end{equation}
As in the logarithmic model, the regular solution exists only if the potential \eqref{2c-10}
opens down, i.e., when $\varkappa > 0$.
The profiles of $\tilde \psi(x)$ and $\tilde \phi(x)$ are presented in Fig. \ref{higgs_functions}. 
For technical reasons in this case an eigenvalue is $\tilde \psi(0)$ not $\tilde E$. In Fig. \ref{higgs_electric_field} the profile of the electric field $\mathcal E$ is shown. In order to show that the electric field asymptotically has Coulomb behaviour we also present the profile $x^2 \mathcal E(x)$ in Fig. \ref{higgs_electric_field}.
From these figures one can see that the qualitative behaviour of the potential $\phi(x)$ and the electric field $\mathcal E(x)$ are the same as for the logarithmic potential. 

\begin{figure}[htbt]
	\begin{center}
		\epsfig{figure=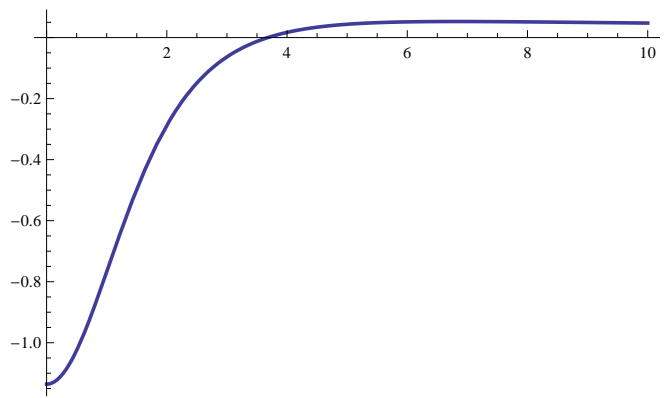,width=  1.0\columnwidth}
	\end{center}
	\caption{Effective potential \eqref{e-effihi} versus $x$ at $\tilde g=1$
	for the quartic model.}
\label{f-effhig}
\end{figure}

The asymptotic behaviour for the functions $\psi$ and $\phi$ at large $x$ is given by 
\begin{eqnarray}
&&
 \tilde \psi(x) \to \psi_\infty \frac{\text{e}^{-x \sqrt{\tilde E+1}}}{x^2}, 
\label{2c-50} \\&&
 \tilde \phi(x) \to -\frac{\tilde q}{x} ,
\label{2c-60}
\end{eqnarray}
where $\psi_{\infty}$ is some constant.
% and $\tilde q$ is the electric charge. 
This still looks very similar to what we had earlier in the logarithmic case,
however if we study the stability of this solution then differences do arise. 
At first,
if one computes the binding energy similarly to \eqref{4c-40} then it turns out to be positive, therefore, the creation of the regular electric charge by coupling electrical field to the quartic scalar one is energetically unfavourable.
Further, if one computes the fictitious-particle potential for this solution, cf. \eqref{e-efflog},
\be\lb{e-effihi}
V_\text{eff} (x)
	= -2 \tilde E \tilde \phi(x) - \tilde \phi^2(x) - 
	\lambda \tilde \psi^2(x) + 1,
\ee
then one finds that it approaches a constant at large $x$ (see also Fig. \ref{f-effhig}). 
The ``particle'' is not, therefore, necessarily localized in a finite region of $x$. 
With respect to the solution itself, this means that the latter can spread or become unstable against small perturbations.

\scn{Conclusion}{sec-con}

The classical model of a spinless electrical charge is described 
as a self-consistent configuration of the EM field interacting with fluctuations of a nontrivial physical vacuum
effectively represented by the logarithmic Bose-Einstein condensate.
We have shown that a regular solution does exist - as opposed
to the case of the EM field propagating in absolutely empty space.
In this regard we recall the state of affairs in quantum mechanics: 
the Dirac/\schrod equation without any external potential has the de Broglie wave solution; 
the Dirac/\schrod equation with the external electrostatic field yields the regular wave
functions (the hydrogen atom being an example),
but the Dirac/\schrod equation coupled to Maxwell equations does not lead to a regular stationary solution. 
The reason is that the Dirac/\schrod equation {\it ab initio} describes a point-like particle 
which might be a good approximation for long-wavelength measurements,
but
in higher-energy and shorter-length regimes this approximation eventually becomes too crude,
since it neglects internal structure and non-zero spatial extent.
Among other things, this leads to the densities of energy and charge becoming infinite
at the particle's position. 
Here we have shown that by
introducing an additional player on scene, the physical vacuum condensate, 
one can obtain a regular solution, thus endowing particles with internal structure
and spatial extent. 
The solution turns out to be stable and energetically favourable.
Using its features, some observational constraints for the parameters of the theory have been derived.
We also specified the conditions under which our model can be (approximately) interpreted
in terms of a scalar particle and those under which it cannot.
 
Further, we have established, both numerically and analytically,
that the mass and spatial extent of a charged particle
emerge due to interaction of the EM field with
vacuum.
It has been demonstrated that the average charge radius becomes non-zero, and
the charge density acquires a Gaussian-like form.
Looking at the form of the analytical solution from Section \ref{sec-appr},
one can infer that it describes the object without border in a classical sense,
therefore, its stability is supported not by surface
tension but
by nonlinear quantum effects in the bulk, similarly to the non-relativistic case \cite{az11}.
Due to  non-singular behaviour of the solution at the origin,
the derivation of self-energy turned out to be entirely divergence-free.

The derived mass formula (\ref{e-massf}) suggested that 
for an  electrically charged massive elementary particle  
there exists not only an antiparticle of the same 
mass
(in the leading-order approximation with respect to the Planck constant,
at least)
but also a neutral particle of related mass.
This might explain, at least qualitatively,
why an electrically charged elementary 
particle is very often accompanied by a single neutral particle 
of a similar kind,
%and only one
but not \textit{vice versa}.
Indeed, such ``mass pairing''
feature has been observed (provided one disregards the influence
of internal degrees of freedom such as spin, isospin, \textit{etc}.)
not only for
elementary particles such as leptons and weak bosons,
but also for stable composite ones such as nucleons
(quarks might not fit this scheme since they are confined inside hadrons).
We presented some arguments for why the rest mass of a neutral partner can sometimes be
so much smaller, yet still non-vanishing, than the mass of the
charged one (leptons), and sometimes they are of the same order of 
magnitude (weak bosons or nucleons).

Finally, we have compared the logarithmic vacuum model  with one based on
a Higgs-type (inverted quartic) potential.
It turns out that 
the corresponding regular solution is unstable and energetically 
unfavourable, in contrast with the logarithmic case.

\begin{acknowledgments}
Both this work and visit of V. D. to the University of Witwatersrand 
were supported by a grant from the National Research Foundation of South Africa.
Authors are grateful to Alexander Avdeenkov for supporting their
visit to the National Institute of Theoretical Physics (Stellenbosch node).
V. D. acknowledges also the 
support from the Research Group Linkage Program of the Alexander von Humboldt Foundation.
K. Z. is grateful to Alessandro Sergi for supporting his
visit to the University of KwaZulu-Natal (Pietermaritzburg campus).
The proofreading
of the manuscript by P. Stannard is greatly acknowledged.
\end{acknowledgments}

\appendix*
\scn{Derivation of approximate solution}{sec-appen}

Here we provide more details regarding derivation of the
approximate analytical solution from Section \ref{sec-appr}.
Assuming \eqref{e-gsmall} and \eqref{e-bc}, we will look for a solution of the system \eqref{2-110} and \eqref{2-120}
in series form
\ba
&&
\tilde\phi = \tilde\Phi_0 (x) + \xi \tilde\Phi_1 (x) + {\cal O} (\xi^2),
\nn\\&&
\tilde\psi = \tilde\Psi_0 (x)
\left(
1 + \xi \tilde\Psi_1 (x) + {\cal O} (\xi^2)
\right)
,
\ea
where $\xi = \tilde g^2 $, and $ \tilde\Phi_i (x)$ and $ \tilde\Psi_k (x)$ are functions
to be determined.
Keeping terms of the order ${\cal O} (\xi)$ and below, equations \eqref{2-110} and \eqref{2-120} can be reduced to the following set of four  differential
equations:
\ba
&&
\tilde\Phi_0'' + \frac{2}{x} \tilde\Phi_0' = 0
,
\\&&
\tilde\Psi_0''
+ \frac{2}{x} \tilde\Psi_0'
+
\left[
\left (
	\tilde E + \tilde \Phi_0  
	\right)^2  
+\ln (\tilde \Psi_0^2)
\right] \tilde \Psi_0
= 0, 
\ \
\\&&
\tilde\Phi_1''
+ \frac{2}{x} \tilde\Phi_1'
-2
\left (
	\tilde E + \tilde \Phi_0  
	\right) 
 \tilde \Psi_0^2
= 0,
\\&&
\tilde\Psi_1''
+
2
\left(
\frac{1}{x} 
+ \frac{\tilde\Psi_0'}{\tilde\Psi_0}
\right) \tilde\Psi_1'
+ \nn\\&&\qquad
+ 2
	(\tilde E + \tilde \Phi_0) \tilde\Phi_1
	+ 2 \tilde\Psi_1
= 0
.
\ea
By solving them we obtain
%of the former three equations satisfying the former of the boundary
\ba
&&
\tilde\phi  = 
\L + 
\xi
\left(
c_2 - \frac{c_1}{x}
\right) - \nn\\&&\qquad
-
\frac{\xi \sqrt{\pi}}{2 x}
(\tilde E + \L) 
\text{e}^{3- (\tilde E + \L)^2}
\text{erf} (x)
 + {\cal O} (\xi^2)
, \quad
\\&&
\tilde\Psi_0  =
\text{e}^{(3- (\tilde E + \L)^2 - x^2)/2}
,
\ea
\ba
&&
\tilde\Psi_1  =
\frac{1}{4}
(\tilde E + \L)^2
\text{e}^{3- (\tilde E + \L)^2 - x^2}
\nn\\&&\qquad
\left[
1
+
\frac{\sqrt{\pi}}{2 x}
(2 x^2 + 1)
\text{e}^{x^2}
\text{erf} (x)
\right]
- \nn\\&&\qquad
-
c_2 (\tilde E + \L)
+  c_3 \left( x - \frac{1}{2 x} \right)
+ \nn\\&&\qquad
+
c_4
\left[
\frac{\sqrt{\pi}}{2 x}
(2 x^2 - 1)
\text{erfi} (x)
-
\text{e}^{x^2}
\right]
,
\ea
where $c_i$ and $\L$ are integration constants
whose values must be fixed by means of the boundary conditions.
Imposing \eqref{e-bc}, one obtains:
$
c_1 = c_3 = c_4 = 0
$ and
$
c_2 = - \L/\xi
$.
Then, after making the energy redefinition $\tilde E + \L \to \tilde E$
(alternatively, one can set $\Lambda = c_2 = 0$ from the beginning),
one eventually arrives at expressions \eqref{e-asolphi} and \eqref{e-asolpsi}.

\def\AnP{Ann. Phys.}
\def\APP{Acta Phys. Polon.}
\def\CJP{Czech. J. Phys.}
\def\CMPh{Commun. Math. Phys.}
\def\CQG {Class. Quantum Grav.}
\def\EPL  {Europhys. Lett.}
\def\IJMP  {Int. J. Mod. Phys.}
\def\JMP{J. Math. Phys.}
\def\JPh{J. Phys.}
\def\FP{Fortschr. Phys.}
\def\GRG {Gen. Relativ. Gravit.}
\def\GC {Gravit. Cosmol.}
\def\LMPh {Lett. Math. Phys.}
\def\MPL  {Mod. Phys. Lett.}
\def\Nat {Nature}
\def\NCim {Nuovo Cimento}
\def\NPh  {Nucl. Phys.}
\def\PhE  {Phys.Essays}
\def\PhL  {Phys. Lett.}
\def\PhR  {Phys. Rev.}
\def\PhRL {Phys. Rev. Lett.}
\def\PhRp {Phys. Rept.}
\def\RMP  {Rev. Mod. Phys.}
\def\TMF {Teor. Mat. Fiz.}
\def\prp {report}
\def\Prp {Report}

\def\jn#1#2#3#4#5{{#1}{#2} {\bf #3}, {#4} {(#5)}} %PRD
%\def\jn#1#2#3#4#5{{#1}{#2} {#3} {(#5)} {#4}}   %PLB style
% #1 tittle  #2 ser  #3 vol  #4 page  #5 year

\def\boo#1#2#3#4#5{{\it #1} ({#2}, {#3}, {#4}){#5}}
%\def\boo#1#2#3#4#5{ #1 ({#2}, {#3}, {#4}){#5}}  %PLB style
% #1 tittle  #2 publisher  #3 place  #4 year  #5 page/, p.789/

%\def\jn#1#2#3#4#5{{#1}{#2} {\bf #3}, {#4} {(#5)}}
% #1 tittle  #2 ser  #3 vol  #4 page  #5 year
%\def\boo#1#2#3#4#5{{\it #1} ({#2}, {#3}, {#4}){#5}}
% #1 tittle  #2 publisher  #3 place  #4 year  #5 page/, p.789/

%\newpage 

%\newpage


\begin{thebibliography}{99}
%\footnotesize


\bibitem{stern52}
A. W. Stern, Science \textbf{116}, 493 (1952).

\bibitem{stern52a}
F. J.  Dyson, Phys. Rev. \textbf{85}, 631 (1952).

\bibitem{stern52b}
L. Van Hove, Physica \textbf{18}, 145 (1952).

\bibitem{stern52c}
P. A. M. Dirac, 
%"The Evolution of the Physicist's Picture of Nature," in 
Scientific American, May 1963, p. 53.

\bibitem{stern52d}
R. P. Feynman, 
%``\textit{
QED - The Strange Theory of Light and Matter,
%}'', 
(Penguin, London, 1990).



\bibitem{ablo}
M. Abraham,
% Prinzipien der Dynamik des Electrons,
Phys. Z. \textbf{4}, 57 (1902).

\bibitem{abloa}
M. Abraham,
Ann. Phys. (Berlin) \textbf{315}, 105 (1902).

\bibitem{ablob}
H. A. Lorentz, 
%"Maxwells elektromagnetische Theorie" 
In: Enzyklop\"adie der Mathematischen Wissenschaften Vol. 5 (Leipzig, Teubner, 1905-1922)
145.


%\cite{Appel:2000th}
\bibitem{Appel:2000th}
R.~Gautreau,
  %``GRAVITATIONAL MODELS OF A LORENTZ EXTENDED ELECTRON,''
  Phys.\ Rev.\  D {\bf 31}, 1860  (1985).
  %%CITATION = PHRVA,D31,1860;%%
	
\bibitem{Appel:2000tha}
  W.~Appel and M.~K.~H.~Kiessling,
  %``Mass and spin renormalization in Lorentz electrodynamics,''
  Annals Phys.\  {\bf 289}, 24 (2001).
  %  [arXiv:math-ph/0009003].
  %%CITATION = APNYA,289,24;%%

%\cite{Dirac:1962iy}
\bibitem{Dirac:1962iy}
  P.~A.~M.~Dirac,
  %``An Extensible model of the electron,''
  Proc.\ Roy.\ Soc.\ Lond.\  A {\bf 268}, 57 (1962).
  %%CITATION = PRSLA,A268,57;%%
	
\bibitem{Dirac:1962iya}
 P.~Gnadig, Z.~Kunszt, P.~Hasenfratz and J.~Kuti,
  %``DIRAC'S EXTENDED ELECTRON MODEL,''
  Annals Phys.\  {\bf 116}, 380 (1978).
  %%CITATION = APNYA,116,380;%%
	
\bibitem{Dirac:1962iyb}
  T.~Fliessbach,
  %``SIMPLIFIED VERSION OF DIRAC'S EXTENSIBLE MODEL OF THE ELECTRON,''
  Am.\ J.\ Phys.\  {\bf 49}, 432 (1981).
  %%CITATION = AJPIA,49,432;%%
	
\bibitem{Dirac:1962iyc}
    A.~O.~Barut and M.~Pavsic,
  %``Dirac's shell model of the electron and the general theory of moving
  %relativistic charged membranes,''
  Phys.\ Lett.\  B {\bf 306}, 49 (1993).
  %%CITATION = PHLTA,B306,49;%%  
  
  
\bibitem{wheelbook}
J. A. Wheeler,
Geometrodynamics (Academic Press, New York, 1962).

%\cite{Born:1934gh}
\bibitem{Born:1934gh}
  M.~Born and L.~Infeld,
  %``Foundations of the new field theory,''
  Proc.\ Roy.\ Soc.\ Lond.\  A {\bf 144}, 425 (1934).
  %%CITATION = PRSLA,A144,425;%%

%\cite{Bronnikov:1979ex}
\bibitem{Bronnikov:1979ex}
  K.~A.~Bronnikov, V.~N.~Melnikov, G.~N.~Shikin and K.~P.~Staniukowicz,
  %``SCALAR, ELECTROMAGNETIC, AND GRAVITATIONAL FIELDS INTERACTION: PARTICLE -
  %LIKE SOLUTIONS,''
  Annals Phys.\  {\bf 118}, 84 (1979).
  %%CITATION = APNYA,118,84;%%
	
\bibitem{Bronnikov:1979exa}
  F.~Finster, J.~Smoller, S.~-T.~Yau,
  %``Particle - like solutions of the Einstein-Dirac-Maxwell equations,''
  Phys.\ Lett.\ A {\bf 259}, 431-436 (1999).
%  [gr-qc/9802012].

\bibitem{Finster:1998ws}  
F.~Finster, J.~Smoller, S.~-T.~Yau,
  %``Particle - like solutions of the Einstein-Dirac equations,''
  Phys.\ Rev.\  D {\bf 59}, 104020 (1999).
%  [gr-qc/9801079].

%\cite{Datzeff:1980tg}
\bibitem{Datzeff:1980tg}
  A.~B.~Datzeff,
  %``ON THE STRUCTURE OF THE ELECTRON,''
  Phys.\ Lett.\  A {\bf 80}, 6 (1980).
  %%CITATION = PHLTA,A80,6;%%
    
%\cite{Bonnor:1989iw}
\bibitem{Bonnor:1989iw}
  W.~B.~Bonnor and F.~I.~Cooperstock,
  %``DOES THE ELECTRON CONTAIN NEGATIVE MASS?,''
  Phys.\ Lett.\  A {\bf 139}, 442 (1989).
  %%CITATION = PHLTA,A139,442;%%  
	
\bibitem{Bonnor:1989iwa}
%Negative energy density and classical electron models  Original Research Article
L. Herrera and V. Varela,
Phys. Lett. A \textbf{189}, 11-14 (1994).


%\cite{Markov:1970hv}
\bibitem{Markov:1970hv}
  M.~A.~Markov,
  %``The closed universe and laws of conservation of electric baryon and lepton
  %charges,''
  Annals Phys.\  {\bf 59}, 109 (1970).
  %%CITATION = APNYA,59,109;%%
%\cite{Markov:1972sc}\bibitem{Markov:1972sc}

\bibitem{Markov:1970hva}
  M.~A.~Markov and V.~P.~Frolov,
  %``On minimal size of particles in general theory of relativity,''
  Teor.\ Mat.\ Fiz.\  {\bf 13}, 41 (1972).
  %%CITATION = TMFZA,13,41;%%
  
%\cite{Zaslavskii:2010yi}
\bibitem{Zaslavskii:2010yi}
E. Ayon-Beato and A. Garcia, 
%Regular Black Hole in General Relativity Coupled to Nonlinear Electrodynamics, 
Phys. Rev. Lett. \textbf{80}, 5056 (1998).

\bibitem{Zaslavskii:2010yia}
K.~A.~Bronnikov,
  %``Comment on 'Regular black hole in general relativity coupled to nonlinear
  %electrodynamics',''
  Phys.\ Rev.\ Lett.\  {\bf 85}, 4641 (2000).
  %%CITATION = PRLTA,85,4641;%%
	
\bibitem{Zaslavskii:2010yib}
S.~Habib Mazharimousavi and M.~Halilsoy,
  %``Black holes and the classical model of a particle in Einstein non-linear
  %electrodynamics theory,''
  Phys.\ Lett.\  B {\bf 678}, 407 (2009).
  %%CITATION = PHLTA,B678,407;%%
	
\bibitem{Zaslavskii:2010yic}
  O.~B.~Zaslavskii,
  %``Zel'dovich states with very small mass and charge in nonlinear
  %electrodynamics coupled to gravity,''
  Grav.\ Cosmol.\  {\bf 16}, 168 (2010) [arXiv:1003.2324].
  %%CITATION = GRCOF,16,168;%%


  
  
\bibitem{bf08}
A. Babin and A. Figotin, arXiv: 0812.2686.




\bibitem{KNold}
B. Carter, 
Phys. Rev. \textbf{174}, 1559 (1968).

\bibitem{KNold1}
W. Israel, 
Phys. Rev. D \textbf{2}, 641 (1970).

\bibitem{KNold2}
G.C. Debney, R.P. Kerr, and A.Schild, 
J. Math. Phys. \textbf{10}, 1842 (1969).

\bibitem{KNold3}
A. Burinskii, 
Sov. Phys. JETP \textbf{39}, 193 (1974).

\bibitem{KNold3a}
A. Burinskii, 
Russian Phys. J. \textbf{17}, 1068 (1974).

\bibitem{KNold3b}
A. Burinskii, 
Phys.\ Rev.\  D {\bf 67}, 124024 (2003).

\bibitem{KNold3c}
A. Burinskii, 
Grav. Cosmol. \textbf{14}, 109 (2008).

\bibitem{KNold4}
D. Ivanenko and A.Ya. Burinskii, 
Russian Phys. J. \textbf{18}, 721 (1975).

\bibitem{KNold5}
C. A. Lopez, 
Phys. Rev. D \textbf{30}, 313 (1984).

\bibitem{KNold5a}
C. A. Lopez, 
Gen. Rel. Grav. 24, 285 (1992).

\bibitem{KNold6}
M. Israelit and N. Rosen, Gen. Rel. Grav. 27, 153 (1995).

\bibitem{KNold7}
%\bibitem{Arcos:2002ip}
  H.~I.~Arcos nad J.~G.~Pereira,
  %``Kerr-Newman solution as a Dirac particle,''
  Gen.\ Rel.\ Grav.\  {\bf 36}, 2441-2464 (2004)
  [hep-th/0210103].
	
\bibitem{KNold8}
%\bibitem{Arcos:2002ip}
H.~I.~Arcos nad J.~G.~Pereira,
%  \bibitem{Arcos:2007id}   H.~I.~Arcos, J.~G.~Pereira,   ``Spacetime: Arena or Reality?,''  
%In: V. Petkov (Ed.), Relativity and the Dimensionality of the World, Fundamental Theories of Physics \textbf{153} (Springer-Verlag, New York, 2007)  
arXiv:0710.0301.
  
%\cite{Burinskii:2011id}
\bibitem{Burinskii:2011id}
  A.~Burinskii,
  %``Gravity versus Quantum theory: Is electron really pointlike?,''
AIP Conf. Proc. \textbf{1424}, 26-32 (2012) [arXiv:1104.0573].
  %%CITATION = ARXIV:1104.0573;%%

%\cite{bs76}
\bibitem{bs76}
K. A. Bronnikov and G. N. Shikin, 
In: Classical and Quantum Theory of Gravity, Trudy IF
(AN BSSR, Minsk, 1976) 88.



%\cite{Latorre:1994cv}
\bibitem{Latorre:1994cv}
  J.~I.~Latorre, P.~Pascual and R.~Tarrach,
  %``Speed of light in nontrivial vacua,''
  Nucl.\ Phys.\  B {\bf 437}, 60 (1995).
 % [arXiv:hep-th/9408016].
  %%CITATION = NUPHA,B437,60;%%


%\cite{Sinha:1976dw}
\bibitem{dir51}
P. A. M. Dirac, 
Nature \textbf{168}, 906 (1951).

\bibitem{dir51a}
P. A. M. Dirac, 
Nature \textbf{169}, 702 (1952).

%\cite{Sinha:1976dw}
\bibitem{Sinha:1976dw}
K.~P.~Sinha, C.~Sivaram and E. C. G.~Sudarshan,
  %``Aether as a superfluid state of particle-antiparticle pairs,''
  Found.\ Phys.\  {\bf 6}, 65 (1976).
  %%CITATION = FNDPA,6,65;%%
%E.~C.~G.~Sudarshan, K.~P.~Sinha and C.~Sivaram,
  %``The Superfluid Vacuum State, Time Varying Cosmological Constant And
  %Nonsingular Cosmological Models,''  Found.\ Phys.\  {\bf 6} (1976) 
	
\bibitem{Sinha:1976dwa}
K.~P.~Sinha, C.~Sivaram and E. C. G.~Sudarshan,
  %``Aether as a superfluid state of particle-antiparticle pairs,''
  Found.\ Phys.\  {\bf 6}, 717 (1976).

\bibitem{Sinha:1976dwb}	
K.~P.~Sinha and E.~C.~G.~Sudarshan,
  %``The Superfluid As A Source Of All Interactions,''
  Found.\ Phys.\  {\bf 8}, 823 (1978).

%\cite{Novello:2002qg}
\bibitem{vol03}
G.~E.~Volovik,
The Universe in a helium droplet,
Int.\ Ser.\ Monogr.\ Phys.\  {\bf 117}, 1-507 (2003).
  %%CITATION = IMPHA,117,1;%% 
  
  
%\cite{Zloshchastiev:2009zw}
\bibitem{Zloshchastiev:2009zw}
K.~G.~Zloshchastiev,
Grav. Cosmol. \textbf{16}, 288 (2010) [arXiv:0906.4282].

\bibitem{Zloshchastiev:2009zwa}
K.~G.~Zloshchastiev,
Phys. Lett. A \textbf{375}, 2305 (2011) [arXiv:1003.0657].

%\cite{Zloshchastiev:2009aw}
\bibitem{Zloshchastiev:2009aw}
K.~G.~Zloshchastiev,
  %``Spontaneous symmetry breaking and mass generation as built-in phenomena in
  %logarithmic nonlinear quantum theory,''
Acta Phys. Polon. B \textbf{42}, 261 (2011) [arXiv:0912.4139].
  %%CITATION = ARXIV:0912.4139;%%
  


\bibitem{ros69}
G. Rosen,
Phys. Rev. \textbf{183} (1969) 1186.  


%\bibitem{kos73} M. D. Kostin, J. Chem. Phys. \textbf{57} (1973) 3589;
%M. D. Kostin, J. Stat. Phys. \textbf{12} (1975) 145.


  
%\cite{BialynickiBirula:1976zp}
\bibitem{BialynickiBirula:1976zp}
I.~Bialynicki-Birula and J.~Mycielski,
  %``Nonlinear Wave Mechanics,''
Annals Phys.\  {\bf 100}, 62 (1976).

\bibitem{BialynickiBirula:1976zpa}
I.~Bialynicki-Birula and J.~Mycielski,
Commun. Math. Phys. \textbf{44}, 129 (1975).
%%CITATION = APNYA,100,62;%%

\bibitem{BialynickiBirula:1976zpb}
I.~Bialynicki-Birula and J.~Mycielski,
%``Gaussons: Solitons Of The Logarithmic Schrodinger Equation. (Talk),''
Phys.\ Scripta {\bf 20}, 539 (1979).
  %%CITATION = PHSTB,20,539;%%
  
  
%\bibitem{lan41} L. D. Landau,
%The theory of superfluidity of helium II. Zh. Eksp. Teor. Fiz. \textbf{11}, 592-614 (1941) [translated in J. Phys. USSR \textbf{5}, 71-90 (1941)].
  
\bibitem{az11}
A.~V.~Avdeenkov and K.~G.~Zloshchastiev,
  %``Quantum Bose liquids with logarithmic nonlinearity: Self-sustainability and
  %emergence of spatial extent,''
J.\ Phys.\ B: At. Mol. Opt. Phys. {\bf 44}, 195303 (2011) [arXiv:1108.0847].

\bibitem{zlo2012}
K.~G.~Zloshchastiev, Eur. Phys. J. B \textbf{85}, 273 (2012) [arXiv:1204.4652].
  
%\cite{Rajaraman:1982is}
\bibitem{Rajaraman:1982is}
R.~Rajaraman,
Solitons and Instantons
(North-Holland, Amsterdam, 1982).
	
\bibitem{Zloshchastiev:2000}
K.~G.~Zloshchastiev,
Phys. Rev. D \textbf{61}, 125017 (2000).

\bibitem{Zloshchastiev:2000a}
K.~G.~Zloshchastiev,
Phys. Lett. B \textbf{519}, 111-120 (2001).  

  



\end{thebibliography}
\end{document}